\documentclass[twocolumn,showpacs,aps,prd]{revtex4-1}
\usepackage{epsfig}
\usepackage{graphicx}
\usepackage{dcolumn}
\usepackage{bm}
\usepackage{ltablex,booktabs}
\usepackage{overpic}
\usepackage{subfigure}
\usepackage{float}
\usepackage{color}
\usepackage{amsmath}
\usepackage{mathcomp}
\usepackage{mathrsfs}
\usepackage{multirow}
\usepackage{rotating}
\usepackage{amssymb}
\usepackage{gensymb}
\usepackage{amsmath}
\usepackage{tabularx}

\begin{document}
\normalsize
\parskip=5pt plus 1pt minus 1pt

\title{\boldmath Search for the decay $D_s^+\to a_0(980)^0e^+\nu_e$}

\author{
\begin{small}
\begin{center}
M.~Ablikim$^{1}$, M.~N.~Achasov$^{10,c}$, P.~Adlarson$^{67}$, S. ~Ahmed$^{15}$, M.~Albrecht$^{4}$, R.~Aliberti$^{28}$, A.~Amoroso$^{66A,66C}$, M.~R.~An$^{32}$, Q.~An$^{63,49}$, X.~H.~Bai$^{57}$
% ...
\\
\vspace{0.2cm}
(BESIII Collaboration)\\
\vspace{0.2cm} {\it
$^{1}$ Institute of High Energy Physics, Beijing 100049, People's Republic of China\\
% ...
$^{4}$ Bochum Ruhr-University, D-44780 Bochum, Germany\\
% ...
$^{10}$ G.I. Budker Institute of Nuclear Physics SB RAS (BINP), Novosibirsk 630090, Russia\\
% ...
$^{15}$ Helmholtz Institute Mainz, Johann-Joachim-Becher-Weg 45, D-55099 Mainz, Germany\\
% ...
$^{28}$ Johannes Gutenberg University of Mainz, Johann-Joachim-Becher-Weg 45, D-55099 Mainz, Germany\\
% ...
$^{32}$ Liaoning Normal University, Dalian 116029, People's Republic of China\\
% ...
$^{49}$ State Key Laboratory of Particle Detection and Electronics, Beijing 100049, Hefei 230026, People's Republic of China\\
% ...
$^{57}$ University of Jinan, Jinan 250022, People's Republic of China\\
% ...
$^{63}$ University of Science and Technology of China, Hefei 230026, People's Republic of China\\
% ...
$^{66}$ University of Turin and INFN, (A)University of Turin, I-10125, Turin, Italy; (B)University of Eastern Piedmont, I-15121, Alessandria, Italy; (C)INFN, I-10125, Turin, Italy\\
$^{67}$ Uppsala University, Box 516, SE-75120 Uppsala, Sweden\\
% ...
$^{c}$ Also at the Novosibirsk State University, Novosibirsk, 630090, Russia\\
% ...
}      
\end{center}
\vspace{0.4cm}
\end{small}
}
\noaffiliation{}

\begin{abstract}
Using 6.32 fb$^{-1}$ of electron-positron collision data recorded by the BESIII 
detector at center-of-mass energies between 4.178 and 4.226~GeV, we present the 
first search for the decay $D_s^+\to a_0(980)^0 e^+\nu_e,\,a_0(980)^0\rightarrow \pi^0\eta$, 
which could proceed via $a_0(980)$-$f_0(980)$ mixing. No significant signal is 
observed. An upper limit of $1.2 \times 10^{-4}$ at the $90\%$ confidence level 
is set on the product of the branching fractions of 
$D_{s}^{+}\to a_0(980)^0 e^+\nu_e$ and $a_0(980)^0\rightarrow \pi^0\eta$ decays.
\end{abstract}
\maketitle

\section{Introduction}

The constituent quark model has been strikingly successful in the past
few decades. The nonets of pseudo-scalar, vector and tensor mesons are
now well identified. On the other hand, the classification of
$J^{\text{PC}}=0^{++}$ scalar mesons still faces difficulty, because
there are more states than predicted by the quark model. Many
theoretical hypotheses have been proposed to explain these extra
states, such as the tetraquark states, two-meson bound states,
molecular-like states, etc.~\cite{PDG}.  More experimental results are
crucial to sort out the interpretations of these states. 
Semileptonic meson decays have a relatively simple decay mechanism and final 
state interactions and can provide a clean probe for studying their hadronic 
part. In particular, semileptonic $D$ meson decays with one scalar meson in the 
final state provide an ideal opportunity to investigate the internal
structures of these light states~\cite{plb-759-501, prd-82-034016}. Example
studies of this type are the semileptonic decays: $D^+\to f_0(500)
e^+\nu_e$, $D^+\to f_0(980) e^+\nu_e$, $D^{0(+)}\to a_0(980)^{-(0)}
e^+\nu_e$, and $D_s^+\to f_0(980) e^+\nu_e$~\cite{prd-80-052007,
  prd-80-052009, prd-92-012009, prl-121-081802, prl-122-062001}.
However, the decay $D_s^+\to a_0(980)^{0} e^+\nu_e$ has not yet been
studied.  

The $D_{s}^+$ direct decay to $a_0(980)^0 e^+\nu_e$ violates isospin
invariance, but it may occur from $D_s^+\to f_0(980) e^+\nu_e$ via
$a_0(980)$-$f_0(980)$ mixing. BESIII has observed
$a_0(980)$-$f_0(980)$ mixing and measured its intensity to be $0.4\%$
in the decays of $J/\psi\to\phi f_0(980)\to\phi a_0(980)^0$ and
$\chi_{cJ}\to f_0(980)\pi^{0}\to
a_0(980)^0\pi^{0}$~\cite{prl-121-022001}. With the product branching fractions~(BF)  
${\cal B}(D_s^+\to
f_0(980) e^+\nu_e)\times{\cal B}(f_0(980)\to\pi^+\pi^-) = (0.13\pm
0.02 \pm 0.01)\times 10^{-2}$~\cite{prd-92-012009} and assuming
$a_0(980)$-$f_0(980)$ mixing effects are the same for $J/\psi$,
$\chi_{cJ}$ and $D_s^+$ decays, one may estimate a BF on the order of $10^{-5}$ for $D_s^+\to a_0(980)^0
e^+\nu_e$. 

A study of the decay $D_s^+\to a_0(980)^0 e^+\nu_e$ could provide
important information on $a_0(980)$-$f_0(980)$ mixing and help to
understand the nature of the scalar meson $a_0(980)$ in the charm
sector. In this paper, the process of $D_s^+\to a_0(980)^0 e^+\nu_e$
with $a_0(980)^0\rightarrow \pi^0\eta$ is studied based on 6.32
fb$^{-1}$ of data recorded by the BESIII detector at center-of-mass
energies ($\sqrt{s}$) between 4.178 and 4.226~GeV. A blind analysis is
performed to avoid a possible bias. Throughout this paper, charge 
conjugate channels are always implied.

\section{DETECTOR and DATA SETS} \label{sec:detector_dataset} Details
about the BESIII detector are described
elsewhere~\cite{Ablikim:2009aa, Ablikim:2019hff}. In short, it is a
magnetic spectrometer located at the Beijing Electron Positron
Collider (BEPCII)~\cite{Yu:IPAC2016-TUYA01}. The cylindrical core of
the BESIII detector consists of a helium-based multilayer drift
chamber~(MDC), a plastic scintillator time-of-flight system~(TOF), and
a CsI(Tl) electromagnetic calorimeter~(EMC), which are all enclosed in
a superconducting solenoidal magnet providing a 1.0~T magnetic
field. The solenoid is supported by an octagonal flux-return yoke with
resistive plate counter muon identifier modules interleaved with
steel. The acceptance of charged particles and photons is 93\% over
$4\pi$ solid angle. The charged-particle momenta resolution at 1.0
GeV/$c$ is $0.5\%$, and the specific energy loss~($dE/dx$) resolution
is $6\%$ for electrons from Bhabha scattering. The EMC measures photon
energies with a resolution of $2.5\%$~($5\%$) at $1$~GeV in the barrel
(end cap) region. The time resolution of the TOF barrel part is 68~ps,
while that of the end cap part is 110~ps. The end cap TOF was upgraded
in 2015 with multi-gap resistive plate chamber technology, providing a
time resolution of 60~ps~\cite{etof}.

Data samples used in this analysis correspond to an integrated
luminosity~($\mathcal{L}_{\rm int}$) of 6.32~fb$^{-1}$ taken in the
range of $\sqrt{s} = $ 4.178 - 4.226~GeV, as listed in
Table~\ref{energe}, and provide a large sample of
$D_{s}^{\pm}$ mesons from $D_{s}^{*\pm}D_s^{\mp}$ events. The cross
section of $D_{s}^{*\pm}D_{s}^{\mp}$ production in $e^{+}e^{-}$
annihilation is about a factor of twenty larger than that of
$D_{s}^{+}D_{s}^{-}$~\cite{DsStrDs} and $D_{s}^{*\pm}$ decays to
$\gamma D_{s}^{\pm}$ with a dominant BF of
$(93.5\pm0.7)$\%~\cite{PDG}. 
\begin{table}[htb]
 \renewcommand\arraystretch{1.25}
 \centering
 \caption{The integrated luminosity $\mathcal{L}_{\rm int}$ and the
   recoil mass $M_{\rm rec}$ requirements for various energies, where
   $M_{\rm rec}$ is defined in Eq.~\ref{eq:mrec}.  The first and
   second uncertainties are statistical and systematic, respectively.}
 \begin{tabular}{ccc}
 \hline 
 $\sqrt{s}$ (GeV) & $\mathcal{L}_{\rm int}$ (pb$^{-1}$) & $M_{\rm rec}$ (GeV/$c^2$)\\
 \hline
  4.178 &$3189.0\pm0.2\pm31.9$&[2.050, 2.180] \\
  4.189 &$526.7\pm0.1\pm2.2$&[2.048, 2.190] \\
  4.199 &$526.0\pm0.1\pm2.1$&[2.046, 2.200] \\
  4.209 &$517.1\pm0.1\pm1.8$&[2.044, 2.210] \\
  4.219 &$514.6\pm0.1\pm1.8$&[2.042, 2.220] \\
  4.226 &$1047.3\pm0.1\pm10.2$&[2.040, 2.220] \\
  \hline
 \end{tabular}
 \label{energe}
\end{table}

Simulated Monte-Carlo~(MC) samples produced with {\sc
  geant4}-based~\cite{GEANT4} software, which includes the geometric
description of the BESIII detector and the detector response, are used
to determine the detection efficiency and to estimate the
background contributions. The simulation includes the beam energy spread and
initial state radiation~(ISR) in the $e^+e^-$ annihilation modeled
with the generator {\sc kkmc}~\cite{KKMC}. Generic MC samples are used
to simulate the background contributions and consist of the production
of $D\bar{D}$ pairs including quantum coherence for all neutral $D$
modes, non-$D\bar{D}$ decays of the $\psi(3770)$, ISR production of
the $J/\psi$ and $\psi(3686)$ states, and continuum processes. The
known decay modes are modeled with {\sc eventgen}~\cite{EVTGEN} using
world averaged BF values~\cite{PDG}, and the remaining
unknown decays from the charmonium states with {\sc
  lundcharm}~\cite{LUNDCHARM}. Final state radiation from charged
final state particles is incorporated with {\sc
  photos}~\cite{PHOTOS}.  The signal detection efficiencies and signal
shapes are obtained from signal MC samples, in which the signal decay
$D_s^+ \to a_{0}(980)^0 e^{+} \nu_{e}$, $a_{0}(980)^0\to \pi^0\eta$,
is simulated using an MC generator where the amplitude of the
$a_{0}(980)^0$ meson follows a theoretical $a_0(980)$-$f_0(980)$
mixing model~\cite{pl-88B-367, prd-82-034016, prd-75-114012, prd-78-074017}. 
This amplitude is given by $A_{\rm mix}=\frac{D_{fa}}{D_{f}D_{a}}$, in which 
$D_{a}$ and $D_{f}$ are the $a_0(980)$ and $f_0(980)$ propagators, 
respectively, and $D_{fa}=\frac{g_{a_0K^+K^-}g_{f_0K^+K^-}}{16\pi}\times
i[\rho_{K^+K^-}(s)-\rho_{K^0\bar{K}^0}(s)]$. Here, $\rho_{K\bar{K}}(s)$ is 
the velocity of the $K$ meson in the rest frame of its mother particle, and 
$g_{a_0K^+K^-}$ and $g_{f_0K^+K^-}$ are coupling constants~\cite{prd-78-074017}.

\section{DATA ANALYSIS}
\label{chap:event_selection}

The signal process $e^{+}e^{-} \to D_{s}^{*+}D_{s}^{-}+c.c.
\to \gamma D_{s}^{+}D_{s}^{-}+c.c$ allows studying
semileptonic $D_{s}^{+}$ decays with a tag
technique~\cite{MarkIII-tag} since only one neutrino escapes
undetected. There are two types of samples used in the tag technique:
single tag (ST) and double tag (DT). In the ST sample, a
$D_{s}^{-}$ meson is reconstructed through a particular hadronic decay
without any requirement on the remaining measured tracks and
EMC showers. In the DT sample, a $D_{s}^{-}$, designated as ``tag'', is
reconstructed through a decay mode first, and then a $D_{s}^{+}$,
designated as the ``signal'', is reconstructed with the remaining tracks
and EMC showers. For one tag mode, the ST yield is given by
\begin{eqnarray} \begin{aligned}
  N_{\text{tag}}^{\text{ST}} = 2N_{D_s^*D_s}\mathcal{B}_{\text{tag}}\epsilon_{\text{tag}}^{\text{ST}}\,, \label{eq-ST}
\label{eq:Ntag}
\end{aligned}
\end{eqnarray}
and the DT yield is given by
\begin{eqnarray}
\begin{aligned}
  N_{\text{tag,sig}}^{\text{DT}}=2N_{D_s^*D_s}\mathcal{B}_{\gamma}\mathcal{B}_{\text{tag}}\mathcal{B}_{\text{sig}}\epsilon_{\text{tag,sig}}^{\text{DT}}\,,
  \label{eq:Nsig} \end{aligned} \end{eqnarray} where $N_{ D_s^*D_s }$
  is the total number of $D_s^{*+}D_s^{-}+c.c.$ pairs produced,
  $\mathcal{B}_{\rm sig (tag)}$ is the BF
  of the signal decay (the tag mode),
  $\mathcal{B}_{\gamma}$ is the BF of $D_s^* \to\gamma D_s$, and
  $\epsilon$ denotes the corresponding reconstruction efficiencies. By
  isolating $\mathcal{B}_{\text{sig}}$, one obtains: \begin{eqnarray}
  \begin{aligned}
  \mathcal{B}_{\text{sig}}=\frac{N_{\text{tag,sig}}^{\text{DT}}\epsilon_{\text{tag}}^{\text{ST}}}{\mathcal{B}_{\gamma}
  N_{\text{tag}}^{\text{ST}}\epsilon_{\text{tag,sig}}^{\text{DT}}},\,
  \label{eq:Bsig} \end{aligned} \end{eqnarray} where the yields
  $N_{\text{tag}}^{\text{ST}}$ and $N_{\text{tag,sig}}^{\text{DT}}$
  can be obtained from data samples, while
  $\epsilon_{\text{tag}}^{\text{ST}}$ and
  $\epsilon_{\text{tag,sig}}^{\text{DT}}$ can be obtained from generic
  and signal MC samples, respectively. The above equations can be
  generalized for multiple tag modes and multiple values of $\sqrt{s}$:
  \begin{eqnarray} \begin{aligned}
  \mathcal{B}_{\text{sig}}=\frac{N_{\text{total,sig}}^{\text{DT}}}{\mathcal{B}_{\gamma}\sum_{\alpha,
  i} N_{\alpha, i}^{\text{ST}}\epsilon^{\text{DT}}_{\alpha,\text{sig},
  i}/\epsilon_{\alpha, i}^{\text{ST}}},\, \label{eq:Bsig-gen}
  \end{aligned} \end{eqnarray} where $\alpha$ represents tag modes,
  $i$ represents different $\sqrt{s}$, and
  $N_{\text{total,sig}}^{\text{DT}}$ is the total signal yield.

The tag candidates are reconstructed with charged $K$ and $\pi$,
$\pi^0$, $\eta^{(\prime)}$, and $K^0_S$ mesons which satisfy the
particle selection detailed below. Twelve tag modes are used and the
requirements on the mass of tagged $D_s^{-}$~($M_{\rm tag}$) are
summarized in Table~\ref{tab:tag-eff}. 

Photons are reconstructed from clusters found in the EMC. The EMC shower
time is required to be within [0, 700]~ns from the event start time in
order to suppress fake photons due to electronic noise or $e^+e^-$
beam background. Photon candidates within $\vert\cos\theta\vert <
0.80$ (barrel) are required to deposit more than 25~MeV of energy, and
those with $0.86<\vert\cos\theta\vert<0.92$ (end cap) must deposit
more than 50~MeV, where $\theta$ is the polar angle with respect to
the $z$ axis, which is the symmetry axis of the MDC. To suppress
Bremsstrahlung photons from charged tracks, the directions of photon
candidates must be at least $10\degree$ away from all charged tracks.
The $\pi^0$ $(\eta)$ candidates are reconstructed through
$\pi^0\rightarrow \gamma\gamma$ ($\eta \rightarrow \gamma\gamma$)
decays, with at least one barrel photon. The diphoton invariant masses
for the identification of $\pi^{0}$ and $\eta$ decays are required to
be in the range $[0.115, 0.150]$~GeV/$c^{2}$ and $[0.490,
0.580]$~GeV/$c^{2}$, respectively. The $\chi^{2}$ of a 1C kinematic
fit constraining $M_{\gamma\gamma}$ to the $\pi^{0}$ or $\eta$ nominal
mass~\cite{PDG} should be less than 30. 

Charged track candidates reconstructed using the information of the
MDC must satisfy $\vert\cos\theta\vert<0.93$ with the closest approach
to the interaction point less than 10~cm in the $z$ direction and less
than 1~cm in the plane perpendicular to $z$. Charged tracks are
identified as pions or kaons with particle
identification~(PID), which is implemented by combining the
information of $dE/dx$ of the MDC and the time-of-flight from the TOF
system. For charged kaon (pion) candidates,
the probability for the kaon (pion) hypothesis is required to be
larger than that for a pion (kaon). For electron identification, the
$dE/dx$, TOF information and EMC measurements are used to construct
likelihoods for electron, pion, and kaon hypotheses (${\cal L}_{e},
{\cal L}_{\pi}$, and ${\cal L}_{K}$).  Electron candidates must
satisfy ${\cal L}_{e}/({\cal L}_{e}+{\cal L}_{\pi}+{\cal L}_{K})>0.7$.
Additionally, the energy measurement using the EMC information of the
electron candidate has to be more than $80\%$ of the track momentum
measured by the MDC ($E/cp>0.8$).

Candidate $K_{S}^{0}$ mesons are reconstructed with pairs of two
oppositely charged tracks, whose distances of closest approach along
$z$ are less than 20~cm. The invariant masses of these
charged track pairs are required to be within $[0.487,
0.511]$ GeV/$c^{2}$.  The $\rho^{0}$ candidates are selected via the
process $\rho^{0} \rightarrow \pi^{+}\pi^{-}$ with an invariant mass
window $[0.570, 0.970]$~GeV/$c^2$. The $\eta^{\prime}$ candidates are
formed from $\pi^{+}\pi^{-}\eta$ and $\gamma\rho^{0}$ combinations
with invariant masses falling within the range of $[0.946, 0.970]$ and
$[0.936, 0.976]$ GeV/$c^{2}$, respectively.

In order to identify the process $e^+e^-\to D^{*\pm}_sD^{\mp}_s$, the
signal windows, listed in Table~\ref{energe}, are applied to the
recoiling mass~($M_{\rm rec}$) of the tag candidate. The definition of
$M_{\rm rec}$ is
\begin{eqnarray}
\begin{aligned}
\frac{1}{c^2}\sqrt{(E_{cm}-\sqrt{c^2|\vec{p}_{\rm
tag}|^2+c^4m^2_{D_s}})^2-c^2|\vec{p}_{cm}-\vec{p}_{\rm tag}|^{2}}\,, 
\label{eq:mrec}
\end{aligned} \end{eqnarray} 
where $(E_{cm}/c,\,\vec{p}_{cm})\equiv p_{cm}$ is the
four-momentum of the $e^+e^-$ center-of-mass system, $(\frac{1}{c}\sqrt{|\vec{p}_{\rm
tag}|^2+m^2_{D_s}},\, \vec{p}_{\rm tag})\equiv p_{\rm tag}$ is the
measured four momentum of the tag candidate, and $m_{D_s}$ is the
nominal $D^{-}_{s}$ mass~\cite{PDG}. If there are multiple candidates
for a tag mode, the one with $M_{\rm rec}$ closest to $D_s^{*\pm}$
mass~\cite{PDG} is chosen.

\begin{figure*}[htp]
\begin{center}
\includegraphics[width=0.80\textwidth]{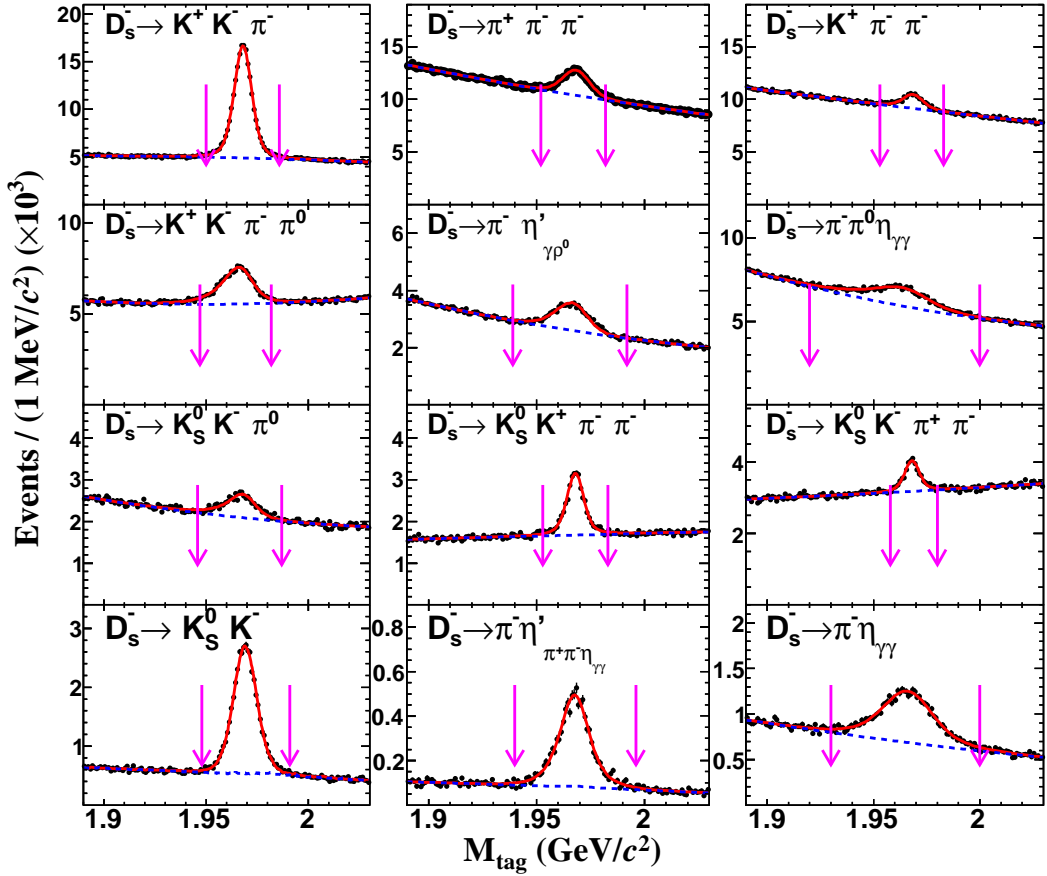}
\caption{Fits to $D^{-}_s$ mass distributions of ST data samples at $\sqrt{s} = 4.178$~GeV. 
         The points with error bars are data, red solid lines are total fits, 
         and blue dashed lines are background. The pairs of pink arrows denote 
         signal regions. MC simulated shapes of $D^{-} \to K_{S}^{0} \pi^-$ and 
         $D_{s}^{-} \to \eta\pi^+\pi^-\pi^-$ decays are added to the background 
         polynomial functions in the fits of $D_{s}^{-} \to K_{S}^{0}K^{-}$ and 
         $D_{s}^{-} \to \pi^{-}\eta^{\prime}$ decays to account for the peaking 
         background, respectively.
         }
\label{fit:Mass-data-Ds_4180}
\end{center}
\end{figure*}

The ST yields for tag modes $N_{\text{tag}}^{\text{ST}}$ are obtained
by fitting the distributions of the tag $D_{s}^{-}$ invariant
mass~($M_{\rm tag}$). Example fits to data samples at 4.178~GeV are
shown in Fig.~\ref{fit:Mass-data-Ds_4180}. The fitting function is an
incoherent sum of the signal and the background contributions. The
description of the signal is based on the MC-simulated shape convolved
with a Gaussian function.  The background is described by a
second-order Chebyshev polynomial function. Based on MC studies, in
all the tag modes, the only significant peaking background is from
$D^{-} \to K_{S}^{0} \pi^-$ and $D_{s}^{-} \to \eta\pi^+\pi^-\pi^-$ decays
faking the $D_{s}^{-} \to K_{S}^{0} K^-$ and $D_{s}^{-} \to
\pi^-\eta^{\prime}$ tag modes, respectively. For these cases, MC
simulated shapes of the two peaking backgrounds are added to the
background polynomial functions. The ST yields of data sample and ST 
efficiencies for tag modes are listed in Table~\ref{tab:tag-eff}.

\begin{table*}[htbp]
 \renewcommand\arraystretch{1.25}
  \caption{Requirements on $M_{\rm tag}$, the ST yields ($N_{\text{tag}}^{\text{ST}}$) 
    and ST efficiencies ($\epsilon_{\text{tag}}^{\text{ST}}$) for
    energy points, (I) $\sqrt{s} = 4.178$~GeV, (II) $4.189$-$4.219$~GeV, and (III) $4.226$~GeV, where the 
    subscripts of $\eta$ and $\eta^\prime$ denote the decay modes used to reconstruct $\eta$ and $\eta^\prime$.
    The efficiencies for the energy points $4.189$-$4.219$~GeV are averaged based on the luminosities.
    The BFs of the sub-particle ($K_{S}^{0}$, $\pi^{0}$, $\eta$ and
    $\eta^{'}$) decays are not included. Uncertainties are statistical only.}\label{tab:tag-eff}
      \begin{center}
      \begin{tabular}{lccccccc}
        \hline
        Tag mode & $M_{\rm tag}$ (GeV/$c^{2}$)  & (I) $N_{\rm tag}^{\rm ST}$  & (I) $\epsilon_{\rm tag}^{\rm ST} (\%)$ & (II) $N_{\rm tag}^{\rm ST}$ & (II) $\epsilon_{\rm tag}^{\rm ST} (\%)$ & (III) $N_{\rm tag}^{\rm ST}$ & (III) $\epsilon_{\rm tag}^{\rm ST} (\%)$\\
        \hline
        $D_{s}^{-} \to K^{+}K^{-}\pi^{-}$            & [1.950, 1.986] & $135859\pm612\phantom{0}$  & $38.96\pm0.03$& $80418\pm503$ & $38.81\pm0.04$ & $28287\pm327\phantom{0}$ & $38.24\pm0.07$\\
        $D_{s}^{-} \to K_{S}^{0}K^{-}$               & [1.948, 1.991] & $31716\pm273$   & $48.89\pm0.06$& $18310\pm227$ & $46.86\pm0.08$ & $6542\pm143$  & $46.36\pm0.15$\\
        $D_{s}^{-} \to \pi^{-}\eta_{\gamma\gamma}$   & [1.930, 2.000] & $18119\pm609$   & $43.07\pm0.15$& $10224\pm458$ & $42.48\pm0.21$ & $3708\pm253$  & $41.75\pm0.40$\\
        $D_{s}^{-} \to \pi^{-}\eta_{\pi^{+}\pi^{-}\eta_{\gamma\gamma}}^{'}$
                                                     & [1.940, 1.996] & $\phantom{0}7799\pm139$    & $19.01\pm0.06$& $\phantom{0}4468\pm111$  & $18.96\pm0.07$ & $1675\pm64\phantom{0}$   & $18.88\pm0.13$\\
        $D_{s}^{-} \to K^{+}K^{-}\pi^{-}\pi^{0}$     & [1.947, 1.982] & $38550\pm772$   & $10.15\pm0.03$& $22945\pm641$ & $10.22\pm0.04$ & $7900\pm437$  & $10.23\pm0.08$\\
        $D_{s}^{-} \to \pi^{-}\pi^{-}\pi^{+}$        & [1.952, 1.982] & $37702\pm852$   & $50.71\pm0.15$& $21517\pm777$ & $49.61\pm0.21$ & $7622\pm542$  & $49.39\pm0.42$\\
        $D_{s}^{-} \to K_{S}^{0}K^{+}\pi^{-}\pi^{-}$ & [1.953, 1.983] & $15637\pm287$   & $21.74\pm0.06$& $\phantom{0}8903\pm233$  & $21.57\pm0.08$ & $3240\pm172$  & $21.28\pm0.15$\\
        $D_{s}^{-} \to \rho^{-}_{\pi^-\pi^0}\eta$    & [1.920, 2.000] & $\phantom{0}41113\pm1324$  & $17.81\pm0.10$& $\phantom{0}25742\pm1203$& $17.89\pm0.14$ & $10729\pm1450$& $17.45\pm0.28$\\
        $D_{s}^{-} \to \pi^{-}\eta^{\prime}_{\gamma\rho^0}$
                                                     & [1.939, 1.992] & $20173\pm603$   & $25.36\pm0.11$& $11364\pm514$ & $25.47\pm0.15$ & $3763\pm727$  & $25.52\pm0.29$\\
        $D_{s}^{-} \to K^{-}\pi^{+}\pi^{-}$          & [1.953, 1.983] & $16939\pm544$   & $45.80\pm0.22$& $10121\pm456$ & $45.38\pm0.30$ & $4918\pm432$  & $44.75\pm0.57$\\
        $D_{s}^{-} \to K_{S}^{0}K^{-}\pi^{0}$        & [1.946, 1.987] & $11260\pm516$   & $15.09\pm0.11$& $\phantom{0}6792\pm469$  & $14.76\pm0.15$ & $2128\pm226$  & $14.84\pm0.27$\\
        $D_{s}^{-} \to K_{S}^{0}K^{-}\pi^{+}\pi^{-}$ & [1.958, 1.980] & $\phantom{0}8013\pm270$    & $20.29\pm0.12$& $\phantom{0}5257\pm289$  & $20.97\pm0.15$ & $1708\pm219$  & $19.45\pm0.30$\\
        \hline
      \end{tabular}
    \end{center}
\end{table*}

\begin{figure*}[htp] \begin{center}
\includegraphics[width=0.40\textwidth]{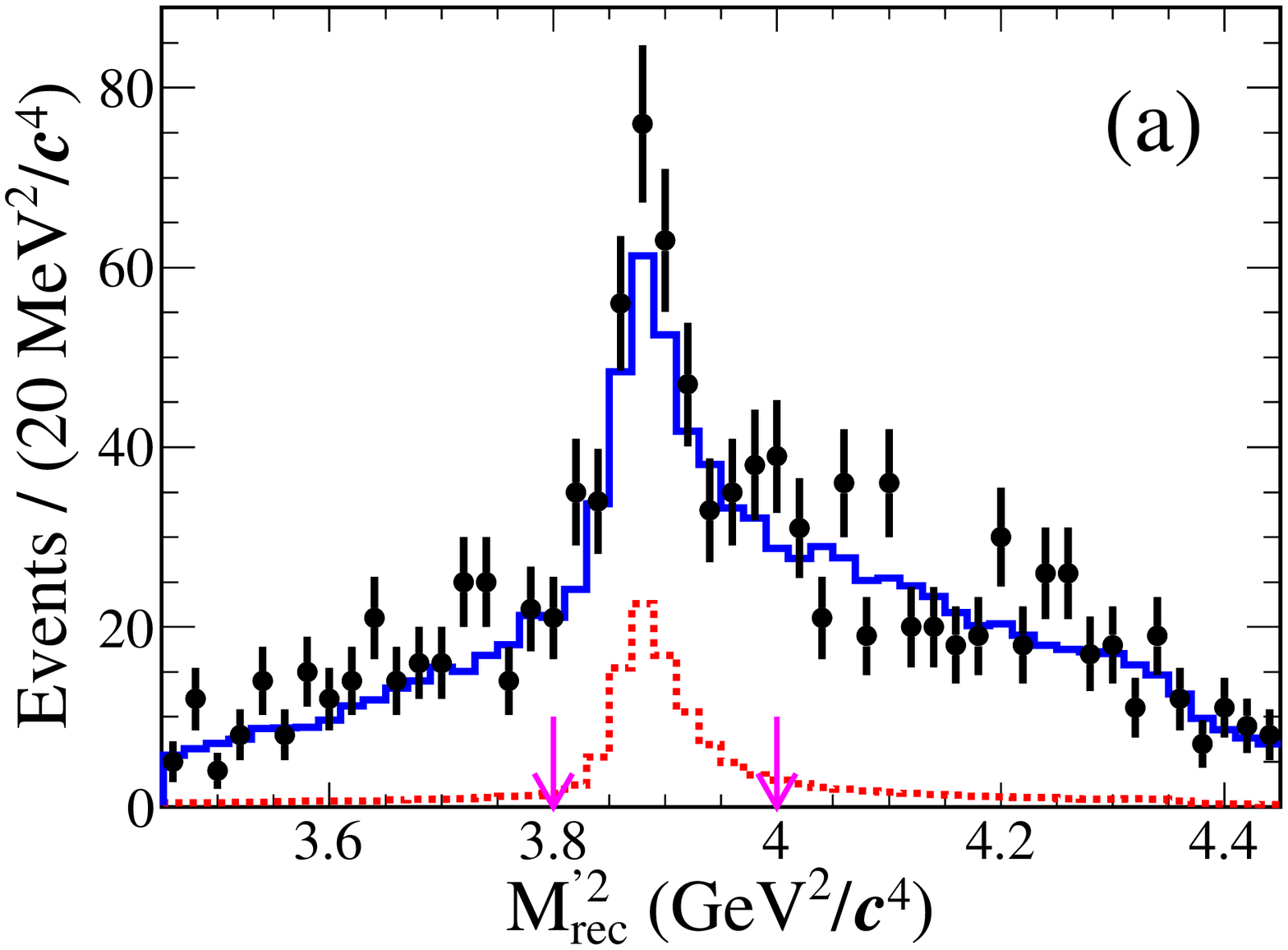}
\includegraphics[width=0.40\textwidth]{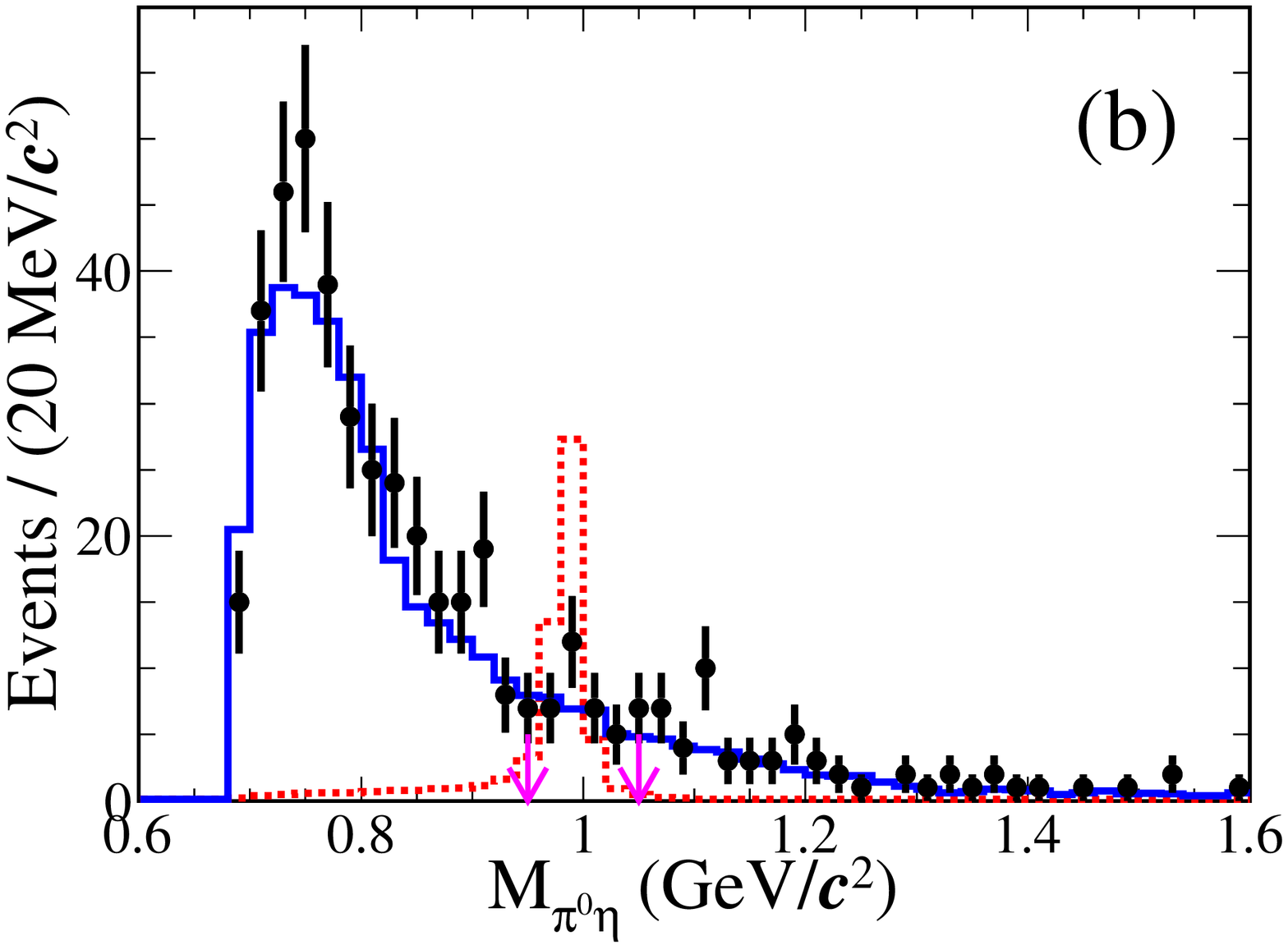}
\caption{(a) $M^{\prime 2}_{\rm rec}$ and (b) $M_{\pi^0\eta}$
distributions of data and MC samples at
         $\sqrt{s} = 4.178$-$4.226$~GeV. The pair of pink arrows denotes the signal windows.
         The points with error bars are data. The blue solid and the red dashed 
         lines are generic and signal MC samples, respectively. 
         The signal MC is normalized arbitrarily for visualization purposes.
         A missing mass cut, $|MM^2|<0.35$~GeV$^{2}$/$c^4$, is applied.
        }
\label{fig:M2rec}
\end{center}
\end{figure*}

After a tag $D_s^-$ is identified, we search for the signal $D_s^+\to 
a_0(980)^0e^+\nu_e,\,a_0(980)^0\rightarrow \pi^0\eta$ recoiling
against the tag by requiring one charged track identified as $e^+$ and
at least five more photons (two for $\pi^0$, two for $\eta$, and one
to reconstruct the transition photon of $D_s^{*\pm}\to\gamma 
D_s^{\pm}$). Events having tracks other than those accounted for in
the tagged $D_s^-$ and the electron are rejected
($N^{\text{extra}}_{\text{char}}=0$).  Kinematic fits are performed on
$e^+e^-\to D_{s}^{*\pm}D_{s}^{\mp}\to \gamma D_{s}^{+}D_{s}^{-}$ with
$D_{s}^{-}$ decays to one of the tag modes and $D_{s}^{+}$ decays to
the signal mode. The
combination with the minimum $\chi^2$ assuming a $D_s^{*+}$ meson
decays to $D_{s}^{+}\gamma$ or a $D_s^{*-}$ meson decays to
$D_{s}^{-}\gamma$ is chosen.  The total four-momentum is constrained
to the four-momentum of $e^+e^-$. Invariant masses of the $D_s^-$ tag,
the $D_s^+$ signal, and the $D_s^*$ are constrained to the
corresponding nominal masses~\cite{PDG}.  Furthermore, it is required
that the maximum energy of photons not used in the DT event selection
($E^{\text{extra}}_{\gamma,\text{max}}$) is less than 0.2~GeV.
Whether the photon forms a $D_s^{*-}$ candidate with the tag $D_s^-$ or a
$D_s^{*+}$ candidate with the signal $D^+_s$, the square of the recoil mass
against the photon and the $D_s^-$ tag ($M^{\prime 2}_{\rm rec}$) should
peak at the nominal $D^{\pm}_s$ meson mass-squared before the kinematics fit
for signal $D^{*\pm}_s D^{\mp}_s$ events.  Therefore, we require $
M^{\prime 2}_{\rm rec}$ to satisfy $3.80 <M^{\prime 2}_{\rm rec}<4.00$~GeV$^{2}$/$c^4$, as shown in
Fig.~\ref{fig:M2rec}(a). To select events from the $a_0(980)^0$ signal
region, the invariant mass of $\pi^0\eta$ ($M_{\pi^0\eta}$) is
required to satisfy $0.95 <M_{\pi^0\eta}<1.05$~GeV/$c^2$, as
shown in Fig.~\ref{fig:M2rec}(b). 

\begin{figure}[htp]
\begin{center}
\includegraphics[width=0.40\textwidth]{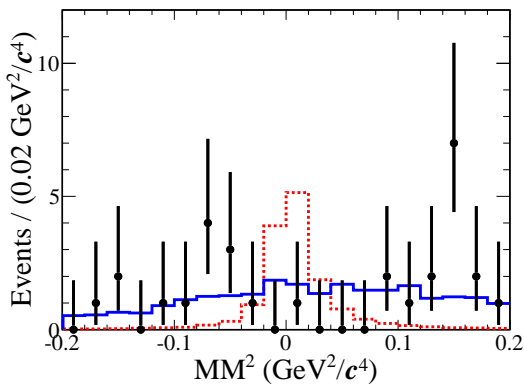}
\caption{ $MM^2$ distributions of data and MC samples at
  $\sqrt{s} = 4.178$-$4.226$~GeV in the signal window. The points with
  error bars are data. The blue solid and the red dashed lines are
  generic and signal MC samples, respectively. The signal MC is
  normalized arbitrarily for visualization purposes.}
\label{fig:MM2_2}
\end{center}
\end{figure}

The missing neutrino is reconstructed by the missing mass squared ($MM^2$), 
defined as

\begin{eqnarray}
\begin{aligned}
MM^2=\frac{1}{c^2}(p_{cm}-p_{\rm tag}-p_{\pi^0}-p_{\eta}-p_{e}-p_{\gamma})^2,\,
\label{def:MM2}
\end{aligned}
\end{eqnarray}
where $p_{i}$ ($i=\pi^0, \eta, e, \gamma$) is the four-momentum of the
daughter particle $i$ on the signal side.  The $MM^2$
distribution of accepted candidate events is shown in
Fig.~\ref{fig:MM2_2}. The DT efficiencies are obtained using the signal
MC samples and listed in Table~\ref{tab:dtagEff2}.
\begin{table*}[htbp]
   \renewcommand\arraystretch{1.25}
  \caption{The DT efficiencies ($\epsilon_{\text{tag,sig}}^{\text{DT}}$) 
    for energy
    points, (I) $\sqrt{s}= 4.178$~GeV, (II) $4.189$-$4.219$~GeV, and (III) $4.226$~GeV. 
    The efficiencies for the energy points $4.189$-$4.219$~GeV are averaged based on the luminosities.
    The BFs of the sub-particle 
    ($K_{S}^{0}$, $\pi^{0}$, $\eta$ and $\eta^{'}$) decays are not included. Uncertainties are statistical only.
  }\label{tab:dtagEff2}
  \begin{center}
    \begin{tabular}{lccc}
      \hline
      Tag mode                                              & (I) $\epsilon_{\text{tag,sig}}^{\text{DT}} (\%)$ & (II)$\epsilon_{\text{tag,sig}}^{\text{DT}} (\%)$&(III)$\epsilon_{\text{tag,sig}}^{\text{DT}} (\%)$\\
      \hline
      $D^-_s\to K^{+}K^{-}\pi^{-}$                          & 4.51 $\pm$ 0.03       & 4.36 $\pm$ 0.02        & 4.17 $\pm$ 0.03\\
      $D^-_s\to K^0_{S}K^{-}$                               & 5.67 $\pm$ 0.08       & 5.42 $\pm$ 0.04        & 5.00 $\pm$ 0.08\\
      $D^-_s\to \pi^{-}\eta$                                & 5.63 $\pm$ 0.09       & 5.46 $\pm$ 0.05        & 4.85 $\pm$ 0.09\\
      $D^-_s\to \pi^{-}\eta^{\prime}_{\pi^{+}\pi^{-}\eta}$  & 2.41 $\pm$ 0.06       & 2.32 $\pm$ 0.03        & 2.16 $\pm$ 0.06\\
      $D^-_s\to K^{+}K^{-}\pi^{-}\pi^{0}$                   & 1.26 $\pm$ 0.02       & 1.25 $\pm$ 0.01        & 1.19 $\pm$ 0.02\\
      $D^-_s\to \pi^{+}\pi^{-}\pi^{-}$                      & 6.25 $\pm$ 0.07       & 5.92 $\pm$ 0.04        & 5.64 $\pm$ 0.07\\
      $D^-_s\to K^0_{S}K^{+}\pi^-\pi^-$                     & 2.55 $\pm$ 0.05       & 2.42 $\pm$ 0.02        & 2.31 $\pm$ 0.04\\
      $D^-_s\to \rho^{-}_{\pi^-\pi^0}\eta$                  & 1.76 $\pm$ 0.02       & 1.68 $\pm$ 0.01        & 1.52 $\pm$ 0.02\\
      $D^-_s\to \pi^{-}\pi^{\prime}_{\gamma\rho^0}$         & 3.66 $\pm$ 0.06       & 3.49 $\pm$ 0.03        & 3.26 $\pm$ 0.06\\
      $D^-_s\to K^{-}\pi^{+}\pi^{-}$                        & 5.48 $\pm$ 0.09       & 5.12 $\pm$ 0.04        & 4.79 $\pm$ 0.09\\
      $D^-_s\to K^{0}_{S}K^{-}\pi^{0}$                      & 2.01 $\pm$ 0.04       & 1.95 $\pm$ 0.02        & 1.84 $\pm$ 0.04\\
      $D^-_s\to K^0_{S}K^{-}\pi^+\pi^-$                     & 2.36 $\pm$ 0.06       & 2.26 $\pm$ 0.03        & 2.22 $\pm$ 0.06\\
      \hline
    \end{tabular}
  \end{center}
\end{table*}
Since no significant signal is observed, an upper limit is
determined. Maximum-likelihood fits to the $MM^2$ distribution
are performed, and likelihoods are determined as a function of assumed
BF.  The signal and the background shapes are modeled by MC-simulated
shapes obtained from the signal MC and the generic MC samples,
respectively.  The likelihood distribution versus BF is shown in
Fig.~\ref{fig:LL_smear_data}.

\section{SYSTEMATIC UNCERTAINTY}
Systematic uncertainties on the BF measurement are summarized in 
Table~\ref{tab:Bf-syst-sum} and the sources are classified
into two types: multiplicative ($\sigma_{\epsilon}$) and additive. 
Note that most systematic 
uncertainties on the tag side cancel due to the tag technique.
\begin{table}[htp]
 \centering
\caption{The multiplicative systematic uncertainties.}
\begin{tabular}{l|c}
\hline
Source                                 & $\sigma_{\epsilon}$ (\%) \\
\hline
${\cal B}(D^*_s\to\gamma D_s)$         & 0.8\\
${\cal B}(\pi^0/\eta\to \gamma\gamma)$ & 0.5\\
$e^+$ Tracking efficiency              & 1.0\\
$e^+$ PID efficiency                   & 1.0\\
$\pi^0/\eta$ reconstruction            & 4.0\\
$\gamma$ reconstruction                & 1.0\\
$E^{\text{extra}}_{\gamma,\text{max}}<0.2$ GeV
                                       & 0.5\\
MC statistics                          & 0.5\\
$N^{\text{extra}}_{\text{char}}=0$     & 0.9\\
Signal model                           & 1.0\\
\hline
Total                                  & 4.7\\
\hline
\end{tabular}
\label{tab:Bf-syst-sum}
\end{table}

Multiplicative uncertainties are from the efficiency determination and
the quoted BFs.  The uncertainty from the BFs of $D^*_s\to \gamma D_s$
and $\pi^0/\eta\to\gamma\gamma$ decays are set to be $0.8\%$ and $0.5\%$,
respectively, according to the world averaged values~\cite{PDG}. The
systematic uncertainties from tracking and PID efficiency of the
$e^{\pm}$, assigned as $1.0\%$, are studied by analyzing radiative
Bhabha events. The systematic uncertainties from reconstruction
efficiencies of neutral particles are determined to be $2\%$ for
$\pi^0$ and $\eta$ by studying a control sample of $\psi(3770)\to
D\bar{D}$ with hadronic $D$ decays, and $1\%$ for $\gamma$ by studying
a control sample of
$J/\psi\to\pi^{+}\pi^{-}\pi^{0}$~\cite{EPJC-76-369,
  CPC-40-113001}. The uncertainties of the
$E^{\text{extra}}_{\gamma,\text{max}}<0.2$ GeV and
$N^{\text{extra}}_{\text{char}} = 0$ requirements are assigned as
$0.5\%$ and $0.9\%$, respectively, by analyzing DT hadronic events,
whereby one $D_s^{\mp}$ decays into one of the tag modes and the other
$D_s^{\pm}$ decays into $K^+K^-\pi^{\pm}$ or $K_{S}K^{\pm}$.  The
parameters of the $a_0(980)$-$f_0(980)$ mixing model in generating the
signal MC samples are varied by $\pm 1\sigma$, and the change of
signal efficiency is assigned as the systematic uncertainty. By adding
these uncertainties in quadrature, the total uncertainty
$\sigma_{\epsilon}$ is estimated to be 4.7\%.

Additive uncertainties affect the signal yield determination, which is
dominated by the imperfect background shape description. The
systematic uncertainty is studied by altering the nominal MC
background shape with two methods. First, alternative MC shapes are
used, where the relative fractions of backgrounds from the major
background source $D_{s}^{+}\to\eta e\nu$, $q\bar{q}$, and
non-$D_{s}^{*+}D_{s}^{-}$ open-charm are varied within their
uncertainties. Second, the background shape is obtained from the
generic MC sample using a kernel estimation method~\cite{kernel}
implemented in RooFit~\cite{RooFit}. The smoothing parameter of
RooKeysPdf is varied to be 0, 1, and 2 to obtain alternative
background shapes. 
\section{Results} 
Since the additive uncertainty is obtained with very
limited sample size, it very likely does not obey a Gaussian
distribution and must be considered conservatively. We repeat the 
maximum-likelihood fits by varying the background shape and take the most 
conservative upper limit among different choices of background shapes.
To incorporate the multiplicative systematic uncertainty in
the calculation of the upper limit, the likelihood distribution 
is smeared by a Gaussian function with a mean of
zero and a width equal to $\sigma_{\epsilon}$ as below~\cite{Stenson, CPC-39-113001}
\begin{equation}
L(n)\propto \int^1_0 L(n\frac{\epsilon}{\epsilon_{0}})exp[\frac{-(\epsilon-\epsilon_0)^2}{2 \sigma_\epsilon^2 }]d\epsilon \,,
\end{equation}
where $L(n)$ is the likelihood distribution as a function of the yield $n$ and 
$\epsilon_0$ is the averaged efficiency. 

\begin{figure}[htp]
  \begin{center}
    \includegraphics[width=0.40\textwidth]{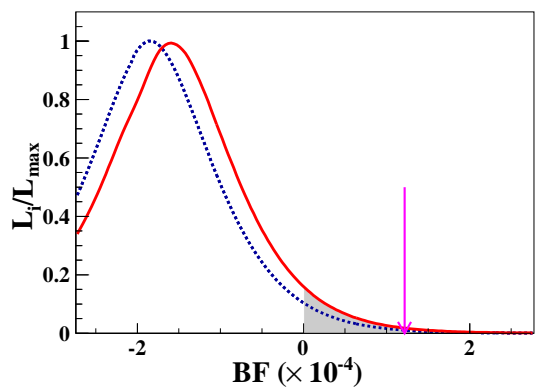}
    \caption{
      Overlays of likelihood distributions versus BF of data samples at 
      $\sqrt{s} = 4.178$-$4.226$~GeV. The results obtained with and without 
      incorporating the systematic uncertainties are shown in red solid and 
      blue dashed curves, respectively. The pink arrow shows the result 
      corresponding to the 90\% confidence level.
    }
    \label{fig:LL_smear_data}
  \end{center}
\end{figure}

The red solid
and blue dashed curves in Fig.~\ref{fig:LL_smear_data} show the
updated and the raw likelihood distributions, respectively. The upper
limit on the BF at the $90\%$ confidence level, obtained by
integrating from zero to $90\%$ of the resulting curve, is ${\cal
B}(D^+_s\to a_0(980)^0 e^+\nu_{e})\times{\cal B}(a_0(980)^0\to
\pi^0\eta)<1.2\times 10^{-4}$.

\section{Conclusion} \label{CONLUSION} Using 6.32~fb$^{-1}$ of data taken
at $\sqrt{s} = $ 4.178-4.226~GeV and recorded by the BESIII detector
at BEPCII, we perform the first search for $D_s^+\to a_0(980)^0
e^+\nu_e$ and obtain an upper limit on ${\cal B}(D^+_s\to a_0(980)^0
e^+\nu_{e})\times{\cal B}(a_0(980)^0\to \pi^0\eta)<1.2\times 10^{-4}$
at the 90\% confidence level.  No obvious isospin violation is
observed.  Comparing to the estimated BF on the order of $10^{-5}$,
this first study of $a_0(980)$-$f_0(980)$ mixing in the charm sector
shows no conflict with the BESIII $a_0(980)$-$f_0(980)$ mixing
measurement results in $J/\psi$ and $\chi_{cJ}$
decays~\cite{prl-121-022001}.

\begin{acknowledgements}
\label{sec:acknowledgement}
\vspace{-0.4cm}
The BESIII collaboration thanks the staff of BEPCII and the IHEP computing center for their strong support. This work is supported in part by National Key Research and Development Program of China under Contracts Nos. 2020YFA0406300, 2020YFA0406400; National Natural Science Foundation of China (NSFC) under Contracts Nos. 11625523, 11635010, 11735014, 11822506, 11835012, 11875054, 11935015, 11935016, 11935018, 11961141012; the Chinese Academy of Sciences (CAS) Large-Scale Scientific Facility Program; Joint Large-Scale Scientific Facility Funds of the NSFC and CAS under Contracts Nos. U1732263, U1832207, U2032104; CAS Key Research Program of Frontier Sciences under Contracts Nos. QYZDJ-SSW-SLH003, QYZDJ-SSW-SLH040; 100 Talents Program of CAS; INPAC and Shanghai Key Laboratory for Particle Physics and Cosmology; ERC under Contract No. 758462; European Union Horizon 2020 research and innovation programme under Contract No. Marie Sklodowska-Curie grant agreement No 894790; German Research Foundation DFG under Contracts Nos. 443159800, Collaborative Research Center CRC 1044, FOR 2359, FOR 2359, GRK 214; Istituto Nazionale di Fisica Nucleare, Italy; Ministry of Development of Turkey under Contract No. DPT2006K-120470; National Science and Technology fund; Olle Engkvist Foundation under Contract No. 200-0605; STFC (United Kingdom); The Knut and Alice Wallenberg Foundation (Sweden) under Contract No. 2016.0157; The Royal Society, UK under Contracts Nos. DH140054, DH160214; The Swedish Research Council; U. S. Department of Energy under Contracts Nos. DE-FG02-05ER41374, DE-SC-0012069.
\end{acknowledgements}



\begin{thebibliography}{99}
\bibitem{PDG} P. A. Zyla {\it et al.} (Particle Data Group), Prog. Theor. Exp. Phys. {\bf 2020}, 083C01 (2020).
\bibitem{plb-759-501} W. Wang, Phys. Lett. B {\bf 759}, 501 (2016).
\bibitem{prd-82-034016} W. Wang and C. D. Lu, Phys. Rev. D {\bf 82}, 034016 (2010).
\bibitem{prd-80-052007} J. Yelton {\it et al.} (CLEO Collaboration), Phys. Rev. D {\bf 80}, 052007 (2009).
\bibitem{prd-80-052009} K. M. Ecklund {\it et al.} (CLEO Collaboration), Phys. Rev. D {\bf 80}, 052009 (2009).
\bibitem{prd-92-012009} J. Hietala {\it et al.}, Phys. Rev. D {\bf 92}, 012009 (2015).
\bibitem{prl-121-081802} M. Ablikim {\it et al.} (BESIII Collaboration), Phys. Rev. Lett. {\bf 121}, 081802 (2018).
\bibitem{prl-122-062001} M. Ablikim {\it et al.} (BESIII Collaboration), Phys. Rev. Lett. {\bf 122}, 062001 (2019).
\bibitem{prl-121-022001} M. Ablikim {\it et al.} (BESIII Collaboration), Phys. Rev. Lett. {\bf 121}, 022001 (2018).
\bibitem{Ablikim:2009aa} M. Ablikim {\it et al.} (BESIII Collaboration), Nucl. Instrum. Methods Phys. Res. Sect. A {\bf 614}, 345 (2010).
\bibitem{Ablikim:2019hff} M. Ablikim {\it et al.} (BESIII Collaboration), Chin. Phys. C {\bf 44}, 040001 (2020).
\bibitem{Yu:IPAC2016-TUYA01} C. H. Yu {\it et al.}, Proceedings of IPAC2016, Busan, Korea, 2016, doi:10.18429/JACoW-IPAC2016-TUYA01.
\bibitem{etof} X.~Li {\it et al.}, Radiat. Detect. Technol. Methods {\bf 1}, 13 (2017);
 Y.~X.~Guo {\it et al.}, Radiat. Detect. Technol. Methods {\bf 1}, 15 (2017);
 P.~Cao {\it et al.}, Nucl.\ Instrum.\ Meth.\ A {\bf 953}, 163053 (2020).
\bibitem{DsStrDs} D. Cronin-Hennessy {\it et al.}  (CLEO Collaboration), Phys. Rev. D \textbf{80}, 072001 (2009).
\bibitem{GEANT4} S. Agostinelli {\it et al.} (GEANT4 Collaboration), Nucl. Instrum. Meth. A {\bf 506}, 250 (2003).
\bibitem{KKMC} S. Jadach {\it et al.}, Phys. Rev. D {\bf 63}, 113009 (2001).
\bibitem{EVTGEN} D. J. Lange, Nucl. Instrum. Meth. A {\bf 462}, 152 (2001); R. G. Ping, Chin. Phys. C {\bf 32}, 599 (2008).
\bibitem{LUNDCHARM} J. C. Chen {\it et al.}, Phys. Rev. D {\bf 62}, 034003 (2000); R. L. Yang {\it et al.}, Chin. Phys. Lett. {\bf 31}, 061301 (2014).
\bibitem{PHOTOS} E. Richter-Was, Phys. Rev. D {\bf 62}, 034003 (2000).
\bibitem{pl-88B-367} N. N. Achasov {\it et al.}, Phys. Lett. {\bf 88B}, 367 (1979).
\bibitem{prd-75-114012} J. J. Wu, Q. Zhao, and B. S. Zou, Phys. Rev. D {\bf 75}, 114012 (2007).
\bibitem{prd-78-074017} J. J. Wu and B. S. Zou, Phys. Rev. D {\bf 78}, 074017 (2008).
\bibitem{MarkIII-tag} J. Adler {\it et al.} (MARK-III Collaboration), Phys. Rev. Lett. {\bf 62}, 1821 (1989).
\bibitem{EPJC-76-369} M. Ablikim {\it et al.} (BESIII Collaboration), Eur. Phys. J. C {\bf 76}, 369 (2016).
\bibitem{CPC-40-113001} M. Ablikim {\it et al.} (BESIII Collaboration), Chin. Phys. C {\bf 40}, 113001 (2016).
\bibitem{kernel} K. S. Cranmer, Comput. Phys. Commun. {\bf 136}, 198 (2001).
\bibitem{RooFit} R. Brun and F. Rademakers, Nucl. Instrum. Methods Phys. Res., Sect. A {\bf 389}, 81 (1997).
\bibitem{Stenson} K. Stenson, arXiv:0605236[physics].
\bibitem{CPC-39-113001} X. X. Liu, X. R. Lyu, and Y. S. Zhu, Chin. Phys. C {\bf 39}, 113001 (2015).
\end{thebibliography}
\end{document}